\documentclass[aip,apl,twocolumn,reprint]{revtex4-1}
\usepackage{graphicx}
\usepackage{dcolumn}
\usepackage{bm}
\usepackage{color}
\usepackage{tabularx}
\usepackage{array}

\usepackage[utf8]{inputenc}
\usepackage[T1]{fontenc}
\usepackage{color,soul}
\usepackage{tabularx}
\usepackage{amsmath}

\makeatother

\begin{document}
\title{Edge-prompted spin waves probed via magneto-optical effect in the infrared}

\author{Junming Wu}
\thanks{These authors contributed equally.}
\affiliation{Department of Physics and Astronomy, University of North Carolina at Chapel Hill, Chapel Hill, NC 27599, USA}

\author{Dinesh Wagle}
\thanks{These authors contributed equally.}
\affiliation{Department of Physics and Astronomy, University of Delaware, Newark, DE 19716, USA}

\author{Yuzan Xiong}
\affiliation{Department of Physics and Astronomy, University of North Carolina at Chapel Hill, Chapel Hill, NC 27599, USA}

\author{Andrew Christy}
\affiliation{Department of Physics and Astronomy, University of North Carolina at Chapel Hill, Chapel Hill, NC 27599, USA}
\affiliation{Department of Chemistry, University of North Carolina at Chapel Hill, Chapel Hill, NC 27599, USA}

\author{Yi Li}
\affiliation{Materials Science Division, Argonne National Laboratory, Argonne, IL 60439, USA}

\author{Shihao Zhou}
\affiliation{Department of Physics and Astronomy, University of North Carolina at Chapel Hill, Chapel Hill, NC 27599, USA}

\author{James F. Cahoon}
\affiliation{Department of Chemistry, University of North Carolina at Chapel Hill, Chapel Hill, NC 27599, USA}

\author{Xufeng Zhang}
\affiliation{Department of Electrical and Computer Engineering, Northeastern University, Boston, MA 02115, USA}
\affiliation{Department of Physics, Northeastern University, Boston, MA 02115, USA}

\author{M. Benjamin Jungleisch}
\thanks{mbj@udel.edu}
\affiliation{Department of Physics and Astronomy, University of Delaware, Newark, DE 19716, USA}

\author{Wei Zhang}
\thanks{zhwei@unc.edu}
\affiliation{Department of Physics and Astronomy, University of North Carolina at Chapel Hill, Chapel Hill, NC 27599, USA}

\begin{abstract}

Recent advances in magnonics highlight the need for employing spin wave characteristics as new state variables, which is to be detected and mapped out with high precision in all-on-chip, small scale devices. Spin wave modes that are prompted by the sample edges in planar structures are of particular fundamental importance to the future design and characterization of
magnonic device functionalities. In this work, we demonstrate an infrared light-based, magneto-optical technique that concurrently functions both as a spectroscopic and an imaging tool. The detected precessional phase contrast can be directly used to construct the map of the spin wave wavefront in the continuous-wave regime of spin-wave propagation and in the stationary state, without needing any optical reference path. We experimentally detect the edge spin wave modes of a Y$_3$Fe$_5$O$_{12}$ film sample, and study the evolution and interference from edge dominated modes to their bulk counterparts. Distinct wavevector groups and propagation characteristics are observed for the edge and bulk spin wave modes. Further, the magnon dispersion relation in the infrared is obtained by Fourier transforming the wavefronts measured in real space, which is a huge advantage to the wavevector-resolved Brillouin light scattering (BLS) technique, where the angle of incidence has to be varied, and is also much less cumbersome than the phase-resolved BLS. Our results demonstrate that the infrared optical strobe light can serve as a versatile platform for magneto-optical probing of magnetization dynamics, with potential implications in investigating hybrid magnonic systems. 

\end{abstract}

\flushbottom
\maketitle

\section{Introduction}

Spin waves (or magnons) are promising chargeless signal carriers for future energy-efficient information systems beyond convention CMOS platforms. Recent rapid developments in magnonic research has already yielded numerous interesting functionalities and promising device applications \cite{flebus20242024,chumak2022advances,xiong2024magnon,xiong2024hybrid,xiong2024combinatorial}. Unlike microwave or light photons, spin waves exist in solid-state magnetic materials and thus, are susceptible to chemical and/or magnetic defects, either structural (e.g. edges, dislocations, and voids) or inherently, magnetic (domain walls, vortices, and skyrmions), often due to the peculiar effective fields surrounding such objects. The sensitive response of spin waves to such 'local-field sources' could inherently be a drawback for any useful applications, however, they may also be leveraged for certain unique functionalities, for example, improving the capability of transmitting information at high frequencies \cite{lara2017information}, channelling and guiding spin waves propagation near a domain wall \cite{yu2021magnetic}, inducing and emitting chiral spin waves via interacting with the gyrotropic mode of a skyrmion \cite{chen2021chiral}, and exciting short-wavelength magnon modes \cite{liu2018long,talapatra2023imaging,xiong2022tunable}.

\begin{figure}[htb]
 \centering
 \includegraphics[width=3.5 in]{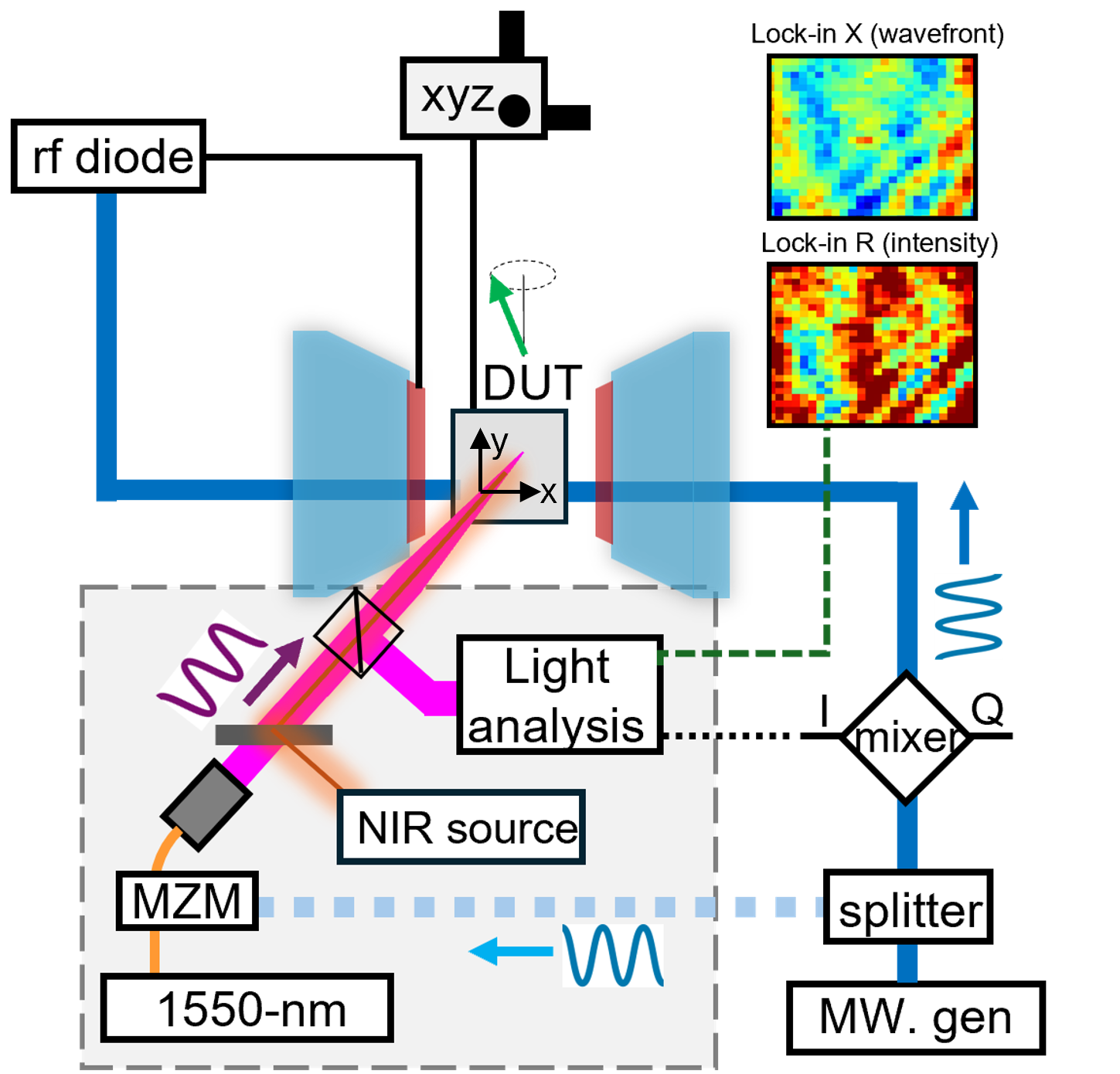}
 \caption{Schematic illustration of the experimental setup. An rf signal, $f_e$, from a single microwave generator (MW. gen) splits into two branches. Along the electrical path, the signal is I/Q-mixed with another intermediate frequency (IF) and is ultimately used to drive the spin dynamics of the DUT, at $f_e+$IF. Along the optical path, the strobe light (intensity-modulated at the original $f_e$ by the MZM) arrives slightly later in the precession cycle and over time, the demodulated magneto-optical response from the DUT traces out complete spin precession cycles. Such a process is realized by analyzing the modulated light polarization, via a polarization beam-splitter followed by a balanced photo-detection module. (light analysis). Further, complementary sample and laser imaging can be achieved by introducing a broadband NIR illumination source and an infrared-sensitive camera. }
 \label{fig:fig1}
\end{figure}

In this sense, spin waves residing at, or interacting with the sample edges in planar structures are of particular fundamental importance to future design and characterization of magnonic devices, in particular for micro- and nanostructured film samples. The surface magnetic charges near the film edges create static demagnetizing field, which is nonuniform and forms wells in the internal magnetic field that confines the low-frequency spin waves, known as the edge modes \cite{kruglyak2006time,jorzick2002spin}. 

To investigate such a topic, magnetic spin wave imaging remains the most powerful technique. In this work, we experimentally investigate the edge spin wave modes of a Y$_3$Fe$_5$O$_{12}$(YIG) film sample and their mode evolution to the bulk counterparts (upon increasing the frequency), by exploiting a stroboscopic light probe operating at the infrared (IR) wavelength (1550-nm) \cite{xiong2024phase}. Different from the existing spin wave imaging techniques such as pump-probes \cite{kirilyuk2010ultrafast,qin2021nanoscale,dreyer2022imaging}, Brillouin light scattering (BLS) \cite{sebastian2015micro,lendinez2023nonlinear,wang2023deeply}, and nitrogen-vacancy(NV) spectroscopy \cite{andrich2017long,zhou2024sensing,bertelli2020magnetic}, the use of the IR strobe light allows direct, concurrent constructions of the spin wave's wavefront and the intensity maps in the stationary state, without needing any additional optical reference paths, such as the pump-line in the time-resolved setup or the complex interferometer in the photon-counting BLS. 

The key principle of the strobe light detection bestows a phase-locked optical oscillation (local probe) to the microwave excitation (global drive) of the sample (or device-under-test, DUT). As depicted in Fig. \ref{fig:fig1}, by simply branching out a portion of the microwave signal for electro-optical modulation, using the Mach-Zehnder modulator (MZM), a strobe light probe by means of magneto-optical Kerr or Faraday effects can be introduced. Such an optical probe is phase-correlated to the microwave drive, thus allowing to accurately trace the spin dynamics referenced to the microwave excitation. The detailed spectroscopic functions and the imaging principles of the technique has been described in earlier reports \cite{xiong2020detecting,xiong2020probing,xiong2022tunable,li2019simultaneous,xiong2024phase}.

\section{Results and Discussions}

\subsection{Strobe light spin-wave spectroscopy}

We first demonstrate the spin wave spectroscopy measurement using the strobe light probe. The sample used for the test is a 23-$\mu$m-thick YIG, cut into a rectangular shape in the dimension of $2 \times 3$ mm$^2$. For reflective optical measurement, we deposited a Ti(5-nm)/Pt(120-nm) metal capping on one YIG side serving as a reflective mirror for the IR light. The sample is flip-mounted atop of a coplanar waveguide (CPW) with the YIG side facing the CPW (for effective excitation). The optical light probes the sample from the backside. The external magnetic field is applied along the horizontal direction ($x$), and the rf field is along $y$. 

\begin{figure}[htb]
 \centering
 \includegraphics[width=3.6 in]{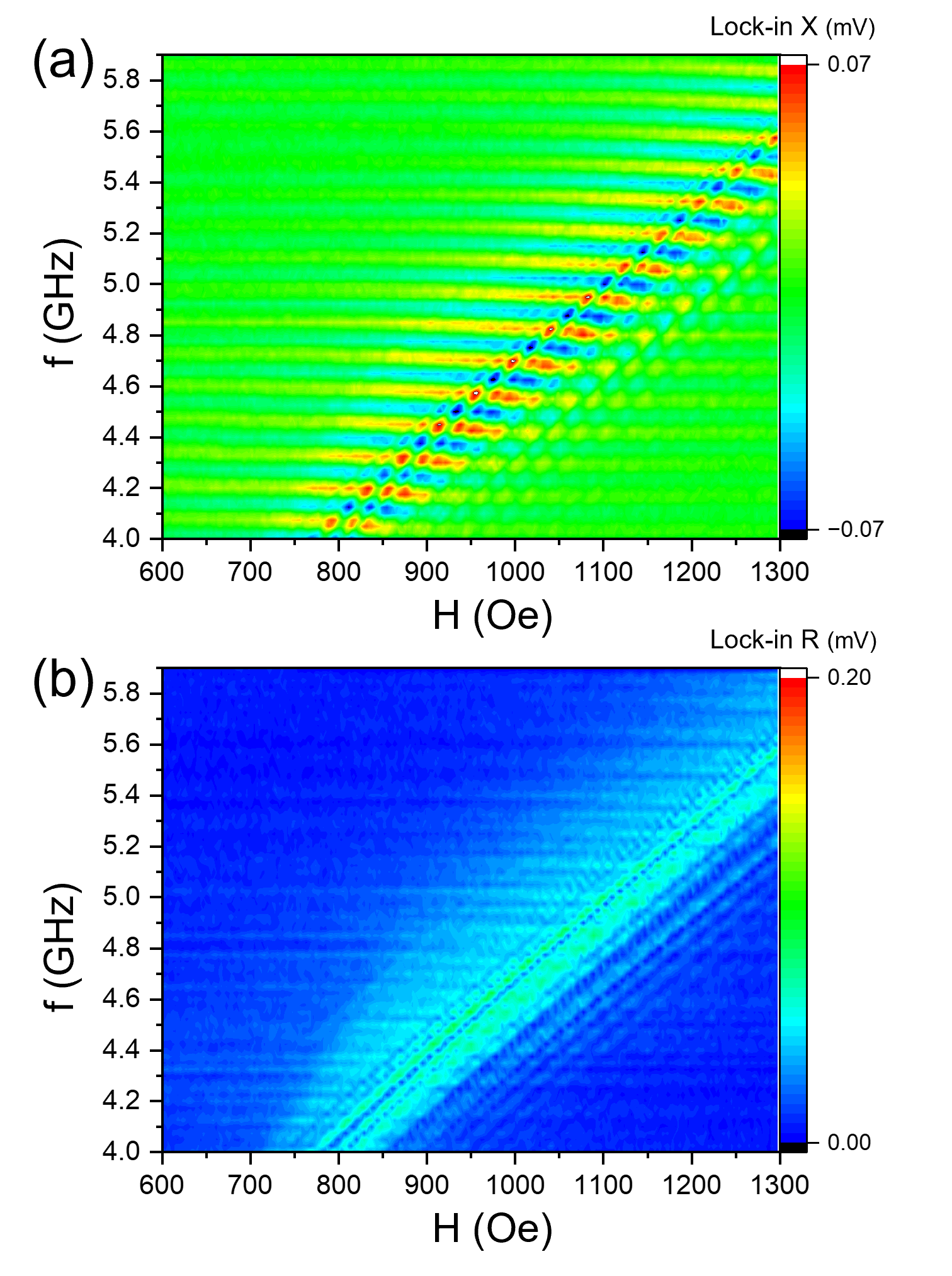}
 \caption{The $f-H$ spectral contour maps obtained by the lock-in amplifier's (a) $X$ channel, and (b) $R$ channel, which are relevant to the construction of the wavefront and intensity spatial mappings, respectively.}
 \label{fig:fig2}
\end{figure}

\begin{figure*}[htb]
 \centering
 \includegraphics[width=7.3 in]{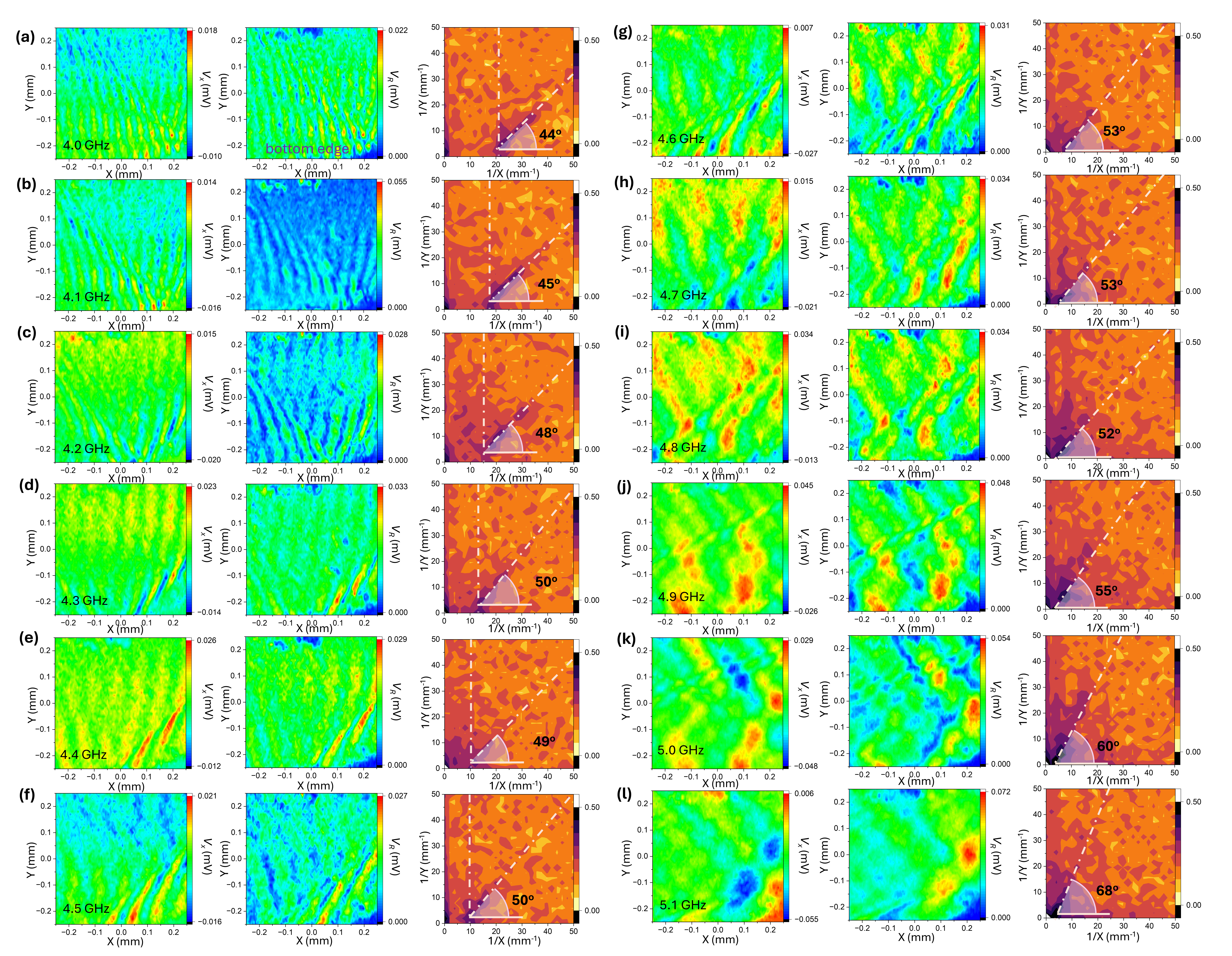} 
 \caption{ (a - l) Scanned 2-D wavefront map (left) and intensity map (middle) as well as the corresponding 2-D FFT map with intensity in log scale (right), at fixed magnetic field $H$ = \textcolor{black}{1225 Oe}, and at selective frequencies (units in GHz): (a) 4.0, (b) 4.1, (c) 4.2, (d) 4.3, (e) 4.4, (f) 4.5, (g) 4.6, (h) 4.7, (i) 4.8, (j) 4.9, (k) 5.0, (l) 5.1. The scanned window, \textcolor{black}{along both horizontal ($X$) and vertical ($Y$),} is $0.5 \times 0.5$ mm$^2$, near the bottom edge of the sample. \textcolor{black}{The tilting angles (representing the k$_x$ / k$_y$ ratio) are labeled for each FFT map.} }  
 \label{fig:fig3}
\end{figure*}

Comprehensive $f - H$ dispersion spectra were measured using the strobe light probe, in which the wavefront signal (lock-in's $X$-channel), Fig. \ref{fig:fig2}(a), and the intensity signal (lock-in's $R$-channel), Fig. \ref{fig:fig2}(b), can be simultaneously obtained. Given the measurement geometry and the spin wave dispersion, we confirm the measured backward volume spin wave (BWVSW) modes in the 23-$\mu$m YIG sample (more details to be discussed later). The BWVSWs are excited through coupling of the \textcolor{black}{out-of-plane (OOP)} component of the alternating rf field with the OOP component of the dynamic magnetization. From the \textcolor{black}{FMR Kittel mode}, we extracted an $M_\textrm{eff}$ value of the YIG to be \textcolor{black}{1740 Oe, in good agreement with typical YIG film samples \cite{xiong2020probing,xiong2022tunable,xiong2024phase}.} For a fixed frequency, the BWVSWs occur at magnetic fields higher than the FMR mode, due to the negative dispersion of BWVSWs \cite{xiong2024phase}. 

\subsection{Imaging spin waves near the sample edge}

Using the strobe light probe, the detected spin precessional phase contrast can be directly used to construct the spinwave wavefront in the stationary state, without needing any additional optical reference paths. By using a long working-distance IR-band objective lens combined with a precise 3-D micro-positioner (x,y,z), we demonstrate the capability of spin-wave spatial mapping in our YIG samples. 

\begin{figure}[htb]
 \centering
 \includegraphics[width=3.2 in]{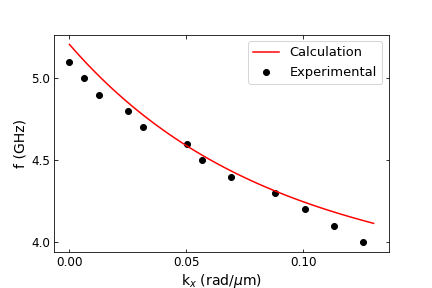}
 \caption{\textcolor{black}{Comparison between experimentally obtained and theoretically expected BWVSW mode dispersion for an external magnetic field of 1225 Oe. k$_x$ values were extracted from Fig.~\ref{fig:fig3} (third columns) at different frequencies along the direction where k$_y = 0$, i.e., where the vertical white dashed, pointed line intersects with the $x$-axis. The red line represents the theoretically calculated dispersion relation for BWVSWs, showing good agreement with the experimental data. }}
 \label{fig:fig4}
\end{figure}

We performed 2-D scan (along $x$-$y$) in the vicinity of the YIG sample edge atop the CPW. Figure \ref{fig:fig3} shows the representative 2-D wavefront maps (left column) and intensity maps (middle column) at a fixed external magnetic field $H=$ \textcolor{black}{1225 Oe}, and at selective frequencies from 4.0 to 5.1 GHz. 

The evolution from the edge-dominated spin waves to the bulk-dominated ones can be clearly identified upon increasing the scanned frequencies. At lower frequencies, e.g. 4.0 -- 4.4 GHz, pronounced spin wave excitations can be observed due to the strong demagnetization field along the sample edge (near the bottom of the scanned area), whereas the bulk excitation is still comparably weak (far from FMR). The excitation is sensitive to the local geometry of the sample edge, in which point defects serve as effective local emitters of spin waves. The propagation direction of the edge spin wave mode is collectively stipulated by the local edge geometry and the position of the point emitter. The propagating wavefront angle at each frequency can be verified from the corresponding FFT spectrum (right column in Fig.\ref{fig:fig3}). A dominating k$_x$ over k$_y$ is observed as a result of the spin wave emission primarily from the bottom edge of the sample. As the waves propagate inside the sample, the wavefronts gradually bend, and a k$_x$ -- k$_y$ tilting angle in the range of 40$^\circ$ -- 50$^\circ$ can be observed with respect to the bottom edge.   

As the frequency increases, e.g. 4.5 -- 4.8 GHz, bulk excitation starts to emerge close to the center of the sample. Such bulk modes are characterized by the well-defined wavevectors along the $x$-direction per our excitation geometry. Notably, at the overlapped regime, the interference between the edge and bulk modes causes a beat pattern formation, e.g., in Fig. \ref{fig:fig3}(g-h). This can be also directly visualized by the distinct wavevector groups from the FFT spectrum at each frequency.    

As the frequency further increases (close to the FMR mode), e.g., at 4.9 -- 5.1 GHz, only the bulk modes become dominant. The propagating wavefront exhibits a higher tilt angle, i.e. close to 55$^\circ$ -- 65$^\circ$, as estimated from the FFT spectra. The bulk spin wave beams induce diffraction patterns that are more symmetric in $x$ and $y$, compared to the interference patterns observed earlier (between edge and bulk modes) \cite{makartsou2024spin,wang2024spin}. Concurrently, due to the longer wavelengths, the distinct wavevector groups also start to smear in the FFT spectra, see Fig. \ref{fig:fig3}(k-l).    

The spatial maps and the FFT results further allow for the extraction and quantification of the wavevectors along the scanned $x$ and $y$ dimensions. In the case of BWVSW mode configuration, the FFT spectra re-affirm the characteristic wavevectors (k$_{SW} \parallel \textbf{H} \parallel \hat{x}$) corresponding to different frequencies. From the $1/X$ - $1/Y$ FFT plots in the right column of Fig.~\ref{fig:fig3}, we extracted the wavevectors of spin waves propagating along the magnetic field direction, i.e., k$_x$ when k$_y$ = 0 at various frequencies. As shown by the black dots in Fig.~\ref{fig:fig4}, k$_x$ increases as the frequency decreases. 
\textcolor{black}{The measured frequency - wavevector dispersion of the spin wave modes was fitted to the theoretical dispersion relation of BWVSWs~\cite{StancilBook} in YIG. We used the effective magnetization $M_{\text{eff}} = 1740$ Oe extracted from the Kittel fit of YIG as input parameter and calculated the dispersion of BWVSW for an applied external magnetic field of 1225 Oe with a gyromagnetic ratio $\gamma = 0.00283$ GHz/Oe. The theoretical dispersion relation, plotted in red, aligns well with the experimental data shown by black symbols, indicating a good agreement between theory and experiment. This close match also re-confirms that the spin wave modes observed in the experiment are predominantly BWVSWs.}

\section{Conclusion}

In summary, the concept of edge spin waves allows to design a broad range of magnonic devices such as splitters, interferometers, or edge wave transistors with augmented characteristics. Laser-based spectroscopy and imaging remains one of the most powerful techniques in such an investigation and alike. Here, we experimentally detect the edge spin wave modes of a YIG film sample and study the evolution and interference from edge dominated modes to their bulk counterparts (upon increasing the frequency), by exploiting a stroboscopic light probe operating at the IR wavelength. Our method shares the similar advantages with conventional lock-in based, field-sweep FMR measurements with high magnetic field resolution and broad dynamic range. The magnon dispersion relations in the infrared are extracted by Fourier transforming the wavefronts measured in real space, an advantage to the conventional BLS techniques where the angle of light incidence has to be varied to acquire any wavevector information. Further, the use of a cw laser and fiber-optics also greatly simplifies the optical layout, allowing the system to be much less susceptible to external noises and more integratable with other optical techniques.

\section*{Acknowledgment}

The experimental work at UNC-CH was supported by the U.S. National Science Foundation (NSF) under Grant No. ECCS-2246254. The analytical work at UD was supported by the U.S. Department of Energy, Office of Basic Energy Sciences, Division of Materials Sciences and Engineering under Award DE-SC0020308. Y.L. acknowledges support from the U.S. DOE, Office of Science, Basic Energy Sciences, Materials Sciences and Engineering Division under contract No. DE-SC0022060. X.Z. acknowledges support from NSF under Grant No. ECCS-2337713.

\bibliography{references}

\end{document}